\documentclass{PoS}

\usepackage[utf8]{inputenc}

\title{Status of Neutrino Astronomy -- a mini-review on neutrino telescopes}

\ShortTitle{Status of Neutrino Astronomy}

\author{\speaker{Alexander Kappes}\\
        Humboldt-Universität zu Berlin / DESY, Institut für Physik, Newtonstr. 15, 12489 Berlin\\
        E-mail: \email{kappes@desy.de}}

\abstract{With the completion of the first cubic-kilometer class neutrino telescopes, IceCube, the race for the discovery of the first cosmic high-energy neutrino sources enters into a new phase. The usage of neutrinos as cosmic messengers has the potential to significantly enhance and extend our knowledge on Galactic and extragalactic sources of the high-energy universe. This article gives a short review on the status of neutrino telescopes and their sensitivities concentrating on point-like sources. It discusses the current upper limits on neutrino emissions and their implications for models of different source classes.}

\FullConference{The 2011 Europhysics Conference on High Energy Physics-HEP 2011,\\
		July 21-27, 2011\\
		Grenoble, Rhône-Alpes France}

\begin{document}

\section{Introduction}
Up to now, information on objects in our galaxy and beyond has nearly exclusively been obtained using electromagnetic waves as cosmic messengers. In addition to this ''electromagnetic'' information, we know from measurements of the cosmic-ray spectrum that there exist sources in the universe which accelerate protons or heavier nuclei up to energies of $\sim 10^{20}$\,eV, $10^7$ times higher than the most energetic man-made accelerator, the LHC at CERN. These highest energies are believed to be reached in extragalactic sources like gamma-ray bursts (GRBs) or active galactic nuclei (AGNs) whereas Galactic sources like supernova remnants (SNRs) or micro-quasars are thought to accelerate particles at least up to energies of $3\times10^{15}$\,eV. However, despite the detailed measurements of the cosmic-ray spectrum and 100 years after their discovery by Victor Hess, we still do not know what the sources of the cosmic rays are as they are deflected in the Galactic and extragalactic magnetic fields and hence have lost all information about their origin when reaching Earth. Only at the highest energies beyond $\sim 10^{19.6}$\,GeV cosmic rays may retain enough directional information to locate their sources.

Alternative messengers for locating the sources of the cosmic rays must have two distinct properties: they have to be electrically neutral and essentially stable. Only two of the known elementary particles meet these requirements: photons and neutrinos. Both particles are inevitably produced when the accelerated protons or nuclei collide with matter or photons inside or near the sources. In these reactions neutral and charged pions are produced which then decay into high-energy photons and neutrinos, respectively. However, only high-energy neutrinos are a smoking-gun evidence for the sources of cosmic rays as TeV photons are also produced in the up-scattering of photons in reactions with accelerated electrons (inverse-Compton scattering).

\section{Neutrino telescopes}
In order to detect the low fluxes of cosmic neutrinos, large volumes of natural transparent media like ice or water have to be instrumented with a three-dimensional array of optical sensors (photomultipliers). Neutrinos are reconstructed by detecting the arrival time and intensity of Cherenkov light from charged secondary particles, which are produced in interactions of the neutrinos with the nuclei in the medium. Two basic event topologies can be distinguished: track-like patterns of detected Cherenkov light (hits) which originate from muons produced in charged-current interactions of muon neutrinos (muon channel); spherical hit patterns which originate from the hadronic cascade at the vertex of neutrino interactions or the electromagnetic cascade of electrons from charged current interactions of electron neutrinos (cascade channel). If the charged current interaction happens inside the detector or in case of charged current tau-neutrino interactions, these two topologies overlap which complicates the reconstruction.

Because of the long lever arm of the muon hit-pattern, the direction of muons can be reconstructed significantly better than that of cascades reaching $\sim 0.1^\circ$ for cubic-kilometer sized detectors at high energies. At the relevant energies, the neutrino is approximately collinear with the muon and, hence, the muon channel is the prime channel for the search for point-like sources of cosmic neutrinos. On the other hand, cascades deposit all of their energy inside the detector and therefore allow for a much better energy reconstruction with a resolution of a few 10\%. 

\begin{figure}
	\center{\includegraphics[width=.65\textwidth]{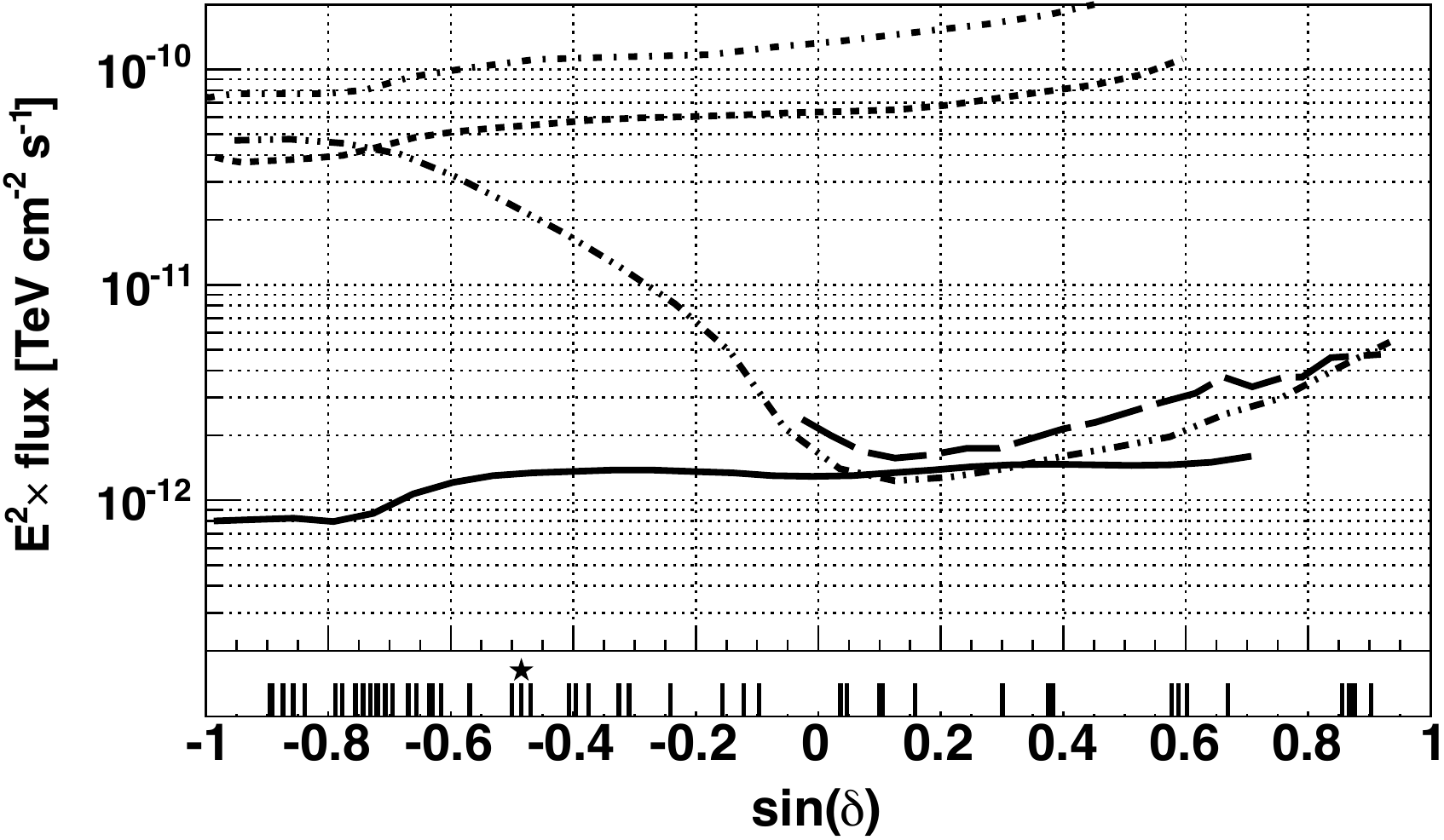} }
	\caption{Upper part: Sensitivities of neutrino telescopes at 90\% CL to a $E^{-2}$ neutrino flux as a function of the source declination: ANTARES (dotted, predicted 1\,yr) \cite{arXiv:1002.0754}, ANTARES 304 days (dashed-dotted) \cite{arXiv:1108.0292}, IceCube 86 strings (dashed, predicted 1\,yr) \cite{apj:732:18}, IceCube 40+59 strings (dashed-dotted-dotted, $375+348$\,d) \cite{proc:icrc11:aguilar:1}, KM3NeT (solid, predicted 1\,yr), \cite{km3net:tdr-short:2010}. Lower part: declination of Galactic objects with observed TeV gamma-ray emission. The position of the Galactic Center is marked with a star.} 
	\label{fig1}
\end{figure}

Currently, the most sensitive neutrino telescope in the Northern Hemisphere is the ANTARES detector, installed off the coast of South France in the Mediterranean Sea at a depth of 2500\,m. In the Southern hemisphere, the currently worldwide largest and hence most sensitive neutrino telescope, IceCube, was completed in Dec.\ 2011. With an instrumented volume of 1\,km$^3$ it is the first cubic-kilometer-class detector and marks a preliminary high point in neutrino telescope development, which have increased their sensitivity by a factor of 1000 in only 15 years. The sensitivities of IceCube and ANTARES to an $E^{-2}$ neutrino flux from a point-like source are shown in Fig.~\ref{fig1}. Source fluxes are expected to lie in the region below $10^{-12}\,\mathrm{TeV}\,\mathrm{cm}^{-2}\,\mathrm{s}^{-1}$. It is apparent that in order to achieve a high-sensitivity coverage over the full sky a cubic-kilometer class detector in the Northern Hemisphere is mandatory which will cover most of the Galactic plane including the Galactic center. Such a detector with an instrumented volume of about 3--6\,km$^3$, KM3NeT \cite{km3net:tdr-short:2010}, is currently in its planning phase. First data can be expected as early as 2014. With IceCube and KM3NeT, neutrino astronomy will hopefully soon become reality.

\section{Status of observations}

\subsection{Galactic sources}
Though it hasn't been possible up to now to identify the sources of Galactic or extragalactic cosmic rays, general considerations allow to limit potential source classes. The local energy density of Galactic cosmic rays is about $10^{-12}\,\mathrm{erg}\,\mathrm{cm}^{-3}$. In order to sustain this, an injection power of about $10^{-26}\,\mathrm{erg}\,\mathrm{cm}^{-3}\,\mathrm{s}^{-1}$ (assuming an escape time of $\sim3\times10^6$\,yr) is needed. Already in 1934, Baade and Zwicky noticed that supernovae could provide this energy. With their rate of about 3 per century and their energy release of $10^{51}$\,erg per explosion, reasonable 10\% of energy has to be converted into the energy of cosmic rays. In addition, in 1949 Fermi devised the basis for the diffusive shock acceleration mechanism which gives a plausible scenario of charged particle acceleration in SNRs. In the meantime, the acceleration of particles to energies of at least 100\,TeV in these shocks has been confirmed by observation of TeV gamma rays. However, these could originate both from inverse Compton scattering of photons from the microwave background radiation off the accelerated electrons (leptonic scenario) or from the decay of neutral pions generated in the collision of cosmic rays with matter fields (hadronic scenario). Though multi wavelength observations in photons can in principle distinguish between these two scenarios, the uncertainties in the measured fluxes and especially the freedom in model building due to the limited information available have and probably also will in the future prevent a definitive answer. On the other hand, the observation of neutrinos from these sources would unambiguously identify them as cosmic-ray accelerators and, hence, bring us a major step forward in our understanding of the high-energy universe.

The direct link between TeV gamma-ray photons and neutrinos through the charged and neutral pion production, which is well known from particle physics, allows for a quite robust prediction of the expected neutrino fluxes provided that the sources are \emph{transparent} and the observed gamma rays originate from \emph{pion decay}. These calculations have been performed by several groups (e.g.\ \cite{apj:656:870}) which all come to the conclusion that in order to detect these fluxes cubic-kilometer sized detectors are necessary. The best environments to search for these neutrino fluxes are star forming regions which provide both a large number of supernovae and, at least as important, large volumes of interstellar gas which acts as a target for the cosmic rays. One of the most promising of these sites is the Cygnus region (distance $\sim 2$\,kpc) which lies in the northern hemisphere and is therefore permanently below the horizon for the IceCube detector. Investigations with IceCube \cite{apj:732:18} have yielded negative results so far. A dedicated stacked search with the half-completed detector for six SNRs seen by Milagro, several of them contained in the Cygnus region, resulted in a moderate p-value of 2\% (posteriori) \cite{apj:732:18}. The follow-up analysis of data with 59 strings installed \cite{proc:icrc11:kurahashi:1} even observed an under-fluctuation of background events. Also, neither for IceCube \cite{apj:732:18,proc:icrc11:aguilar:1,arxiv:1108.3023} nor for ANTARES \cite{arXiv:1108.0292}, do the results for other Galactic sources, both steady and variable, currently show an indication of a deviation from the background.

Thus, the search for Galactic high-energy neutrino sources has been unsuccessful up to now, but this had to be expected from the predicted fluxes. IceCube data analyzed so far has been taken with only a partially completed detector. The investigation of the data set from the nearly completed detector (79 strings installed) is currently underway and the data taking with the full 86 strings is running since April 2011. In the upcoming years, IceCube will further significantly improve its sensitivity and enter into the region of predicted fluxes. The detection of the first Galactic neutrino source might therefore be just around the corner.

\subsection{Extragalactic sources}
As in the case of Galactic sources, general requirements limit the number of potential sources. First of all, the candidate sources must be capable to accelerate particles up to the observed energies of $10^{20}$\,eV. Basic source parameters to reach these energies are the size of the source and the strength of its magnetic field. Both together must be such that the particles are contained long enough in the acceleration region to reach the energies. In addition, the sources must be powerful enough to sustain the energy density in extragalactic cosmic rays of about $3\times10^{-19}\,\mathrm{erg}\,\mathrm{cm}^{-3}$ which is equivalent to $\sim 8\times10^{44}\,\mathrm{erg}\,\mathrm{Mpc}^{-3}\,\mathrm{yr}^{-1}$. It turns out that currently there are only two good candidate source classes: AGNs and GRBs. 

In 2008, the Pierre Auger experiment reported a weak 1\% correlation of their cosmic-ray events above $6\times10^{19}$\,eV with nearby AGNs \cite{sc:318:938}. However, this significance has decreased since then \cite{app:34:314}. In addition, the result is in conflict with their own composition measurements, which prefer a heavy composition (the correlation requires the particles to be protons as heavier nuclei would be deflected too strongly in the magnetic fields), but which have their own problems due to their strong simulation dependency. Therefore, the situation will likely remain unclear for a while. Neutrino predictions for these sources are difficult. Gamma-ray photons are potentially absorbed already inside the source and cascade down to lower energies. As in the case for Galactic sources, both leptonic and hadronic models exist. The predicted neutrino fluxes in the hadronic models are rather low, even for cubic-kilometer sized detectors. But as these sources sometimes exhibit strong and long flaring periods, a detection with IceCube is not impossible. Stacking of several sources will further improve the chances for discovery.

\begin{figure}
	\center{\includegraphics[width=.6\textwidth, height=5.5cm]{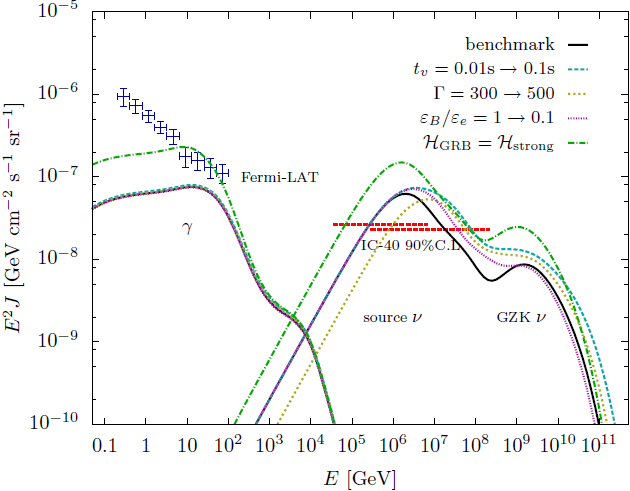}}
	\caption{Neutrino and photon fluxes as a function of energy from GRBs within the fireball model for the case that GRBs are the major source of the UHECRs. The variables are: $t_v=$ variability; $\Gamma =$ bulk Lorentz factor; $\epsilon_B/\epsilon_e=$ ratio of energy in magnetic field and electrons; ${\cal H}_\mathrm{GRB}=$ source evolution. The benchmark scenario has standard values of $t_v=0.01$\,s, $\Gamma=300$, $\epsilon_B/\epsilon_e=1$ and ${\cal H}_\mathrm{GRB}$ equal to the evolution of the star formation rate. Taken from a talk by F. Halzen at Nusky 2011. For more information on the model see \cite{app:35:87}.} 
	\label{fig2}
\end{figure}

The situation is somewhat different for GRBs. Within the widely used fireball model of GRBs, the photons of the gamma-ray flash in the keV--MeV range, observed by satellites like Swift and Fermi, act as target for the accelerated protons. Through the $\Delta$ resonance charged pions are produced which then decay into neutrinos. The resulting neutrons escape the fireball and decay into protons which are then observed as the ultra-high energy cosmic rays (UHECRs). The caveat is that apart from measured parameters like the photon spectrum this model also contains several variables like the $\Gamma$ factor of the ejected material which in most or all cases have to be chosen from theoretical considerations. In the case of $\Gamma$ for example, the standard value in these calculations is $\sim 300$ whereas observations indicate that at least in some cases, though unlikely in all, it can be as high as 1000. Nevertheless, these parameters can only be varied in a reasonable range and not all combinations of variations are in agreement with observations. Figure~\ref{fig2} shows the expected neutrino flux as a function of energy for the case that GRBs are the major sources of the UHECRs. The model parameters where chosen such that all parameters are within a reasonable theoretical range and that the high-energy gamma-rays, which are produced through the decay of neutral pions and afterwards cascade down to lower energies while moving through the interstellar photon fields, are below the diffuse extragalactic flux of GeV photons observed by Fermi. As can be seen, the upper limits on the diffuse neutrino flux from IceCube already exclude all investigated models and hence start to question the role of GRBs as major sources of UHECRs. A significant gain in sensitivity is reached with dedicated searches for neutrinos from  GRBs which take into account the time and position measured by satellites. Also these searches have yielded negative results so far \cite{prl:106:141101} and they have the potential to eventually exclude GRBs as major sources of UHECRs.

\section{Conclusions}
Up to now, all searches for extraterrestrial high-energy neutrinos have yielded negative results. However, with neutrino telescopes of the cubic-kilometer class we are now entering into a sensitivity region where first discoveries are around the corner. Where exactly these discoveries will be made is hard to forsee as the predicted fluxes are at the edge of detectability and the uncertainties on the neutrino fluxes are still large. Therefore, it is important not to focus too much on a particular scenario but stay open-minded and also expect the first detection from an unexpected direction. 

\providecommand{\href}[2]{#2}\begingroup\raggedright\endgroup



\end{document}